\documentclass[preprint,lineno]{biometrika}

\usepackage{amsmath}

\usepackage{times}
\usepackage{bm}
\usepackage{natbib}
\usepackage{amsmath}
\usepackage{setspace,graphicx}     
\usepackage{amsfonts}
\usepackage{latexsym}
\usepackage{upgreek}
\usepackage{amssymb}
\usepackage{setspace}
\usepackage{fancyhdr}
\usepackage{mathrsfs}
\usepackage{framed}
\usepackage{bm}
\usepackage{perpage} 
\usepackage{comment}
\usepackage{bbm}
\usepackage{enumerate}
\usepackage[T1]{fontenc}
\usepackage{fixltx2e}
\usepackage{mathrsfs}
\usepackage{color}

\usepackage[plain,noend]{algorithm2e}

\newcommand{\Var}{\mbox{Var}}

\newcommand{\R}{\mathbb{R}}
\newcommand{\p}{\mathbb{P}}

\newcommand{\hiddensection}[1]{
\stepcounter{section}
\section*{}}

\newcommand{\n}{\mathcal{N}}

\newcommand{\Q}{\mathbb{Q}}

\newcommand{\F}{\mathscr{F}}
\newcommand{\X}{\mathcal{X}}

\newcommand{\envelope}{(\raisebox{-.5pt}{\scalebox{1.45}{\Letter}}\kern-1.7pt)}

\newcommand{\dkl}{D_{\mbox {\tiny{\rm KL}}}}
\newcommand{\dtv}{D_{\mbox {\tiny{\rm TV}}}}
\newcommand{\dchi}{D_{\mbox {\tiny{$ \chi^2$}}}}
\newcommand{\df}{D_{\mbox {\scriptsize{$ f$}}}}
\newcommand{\dhell}{D_{\mbox {\tiny{\rm Hell}}}}

\newcommand{\nkl}{N_{\mbox {\tiny{\rm KL}}}}
\newcommand{\ntv}{N_{\mbox {\tiny{\rm TV}}}}
\newcommand{\nchi}{N_{\mbox {\tiny{$ \chi^2$}}}}
\newcommand{\nhell}{N_{\mbox {\tiny{\rm Hell}}}}

\newcommand{\fkl}{{\cal F}_{\mbox {\tiny{\rm KL}}}}
\newcommand{\fchi}{{\cal F}_{\mbox {\tiny{$ \chi^2$}}}}

\newcommand{\esskl}{{\rm ESS}_{\mbox {\tiny{\rm KL}}}}
\newcommand{\esschi}{{\rm ESS}_{\mbox {\tiny{$ \chi^2$}}}}

\definecolor{darkgreen}{rgb}{0,.6,0}
\definecolor{darkyellow}{rgb}{1,.5,.3}

\newcommand{\si}[1]{\lambda}

\makeatletter
\renewcommand{\algocf@captiontext}[2]{#1\algocf@typo. \AlCapFnt{}#2} 
\def\@algocf@capt@plain{top}
\renewcommand{\algocf@makecaption}[2]{%
  \addtolength{\hsize}{\algomargin}%
  \sbox\@tempboxa{\algocf@captiontext{#1}{#2}}%
  \ifdim\wd\@tempboxa >\hsize
    \hskip .5\algomargin%
    \parbox[t]{\hsize}{\algocf@captiontext{#1}{#2}}
  \else%
    \global\@minipagefalse%
    \hbox to\hsize{\box\@tempboxa}
  \fi%
  \addtolength{\hsize}{-\algomargin}%
}
\makeatother

\begin{document}

\jname{---}
\jyear{2016}
\jvol{---}
\jnum{---}


\markboth{D. Sanz-Alonso}{Importance Sampling and Necessary Sample Size}

\nolinenumbers
\title{ Importance Sampling and Necessary Sample Size: an Information Theory Approach}

\author{D. SANZ-ALONSO}
\affil{Division of Applied Mathematics, Brown University\\Providence, Rhode Island, 02906, US \email{daniel\_  sanz-alonso1@brown.edu}}

\maketitle

\begin{abstract}
Importance sampling approximates expectations with respect to a {\em{target}} measure by using samples from a {\em{proposal}} measure. The performance of the method over large classes of test functions depends heavily on the closeness between both measures. We derive a general bound that {\em needs} to hold for importance sampling to be successful, and relates the $f$-divergence between the target and the proposal to the sample size. The bound is deduced from a new and simple information theory paradigm for the study of importance sampling. As examples of the general theory we give necessary conditions on the sample size in terms of the Kullback-Leibler and $\chi^2$ divergences, and  the total variation and Hellinger distances. Our approach is non-asymptotic, and its generality allows to tell apart the relative merits of these metrics. Unsurprisingly, the non-symmetric divergences give sharper bounds than total variation or Hellinger. Our results extend existing necessary conditions ---and complement sufficient ones--- on the sample size required for importance sampling.
\end{abstract}

\begin{keywords}
$f$-divergence; Importance Sampling; Information Theory; Sample Size.
\end{keywords}

\section{Introduction}
Let $\p$ and $\Q$ be, throughout, two probability measures on a measurable space $(\X, \F),$ with $\p$ absolutely continuous with respect to $\Q.$ Importance sampling  is a Monte Carlo technique that approximates expectations with respect to the \emph{target} $\p$ by using samples from the \emph{proposal} $\Q.$   Our aim is to introduce a simple information theory paradigm to determine situations where this method cannot be successful. The results are non-asymptotic, and are based on information barriers on $f$-divergences.

In what follows it is best to view importance sampling as a way to approximate the target $\p$ by a weighted empirical measure 
\begin{equation}\label{eq:weightedmeasure}
\pi^N := \sum_{n=1}^N w^n \delta_{v^n},
\end{equation}
 where $N$ is the number of samples $v^n$ drawn from $\Q$. 
We now present heuristically the main idea. Let $\Q_{\text{\tiny MC}}^N := 
(1/N) \sum \delta_{v^n} $ denote the standard Monte Carlo approximation of $\Q$.  Importance sampling replaces the uniform weights ${\bf u} = \{1/N, \ldots, 1/N\}$ associated with the particles ${\bf v}^N = \{v^n\}_{n=1}^N$ by non-uniform weights ${\bf w}^N=\{w^n\}_{n=1}^N.$ The hope is that then \eqref{eq:weightedmeasure} approximates $\p$ rather than $\Q.$ Let $D$ be some notion ---to be made precise--- of distance that allows to assess the closeness of  measures in $(\X,\F),$ and also of probability vectors in $[0,1]^N.$ Suppose that the bound $D({\bf w}^N , {\bf u}) \le U(N)$ holds for any choice of non-negative weights ${\bf w}^N$ with $\sum w^n=1$ and a given function $U.$ If we can guarantee ---under conditions ensuring the success of standard Monte Carlo and importance sampling--- that  $D(\p,\Q)$ is close to $D(\pi^N, \Q_{\text{\tiny MC}}^N)$, then we would like to conclude that
\begin{equation*}\label{eq:informationbarrier}
D(\p,\Q) \approx D(\pi^N, \Q_{\text{\tiny MC}}^N) = D({\bf w}^N , {\bf u})   \le U(N).
\end{equation*}
There is an information barrier: if $N$ is such that $U(N)<D(\p,\Q),$ then $N$ samples from $\Q$ do not contain enough information on the model $\p$ for importance sampling to be successful. This heuristic is made rigorous in our main results, Theorems \ref{theorem} and \ref{theorem2}  below. Note that  $\pi^N$ (respectively $\Q_{\text{\tiny MC}}^N$)  will never be close to $\p$ (respectively $\Q)$ in the metrics considered below, since the former is atomic and the latter, in general, is not. However, it is still possible to guarantee  that $D(\p,\Q) \approx D(\pi^N, \Q_{\text{\tiny MC}}^N)$ under appropriate performance conditions on Monte Carlo and importance sampling.

The first step in formalizing our argument is to agree on a metric to assess the closeness of measures. This is a major point, since this choice typically impacts the convergence ---or the rate of convergence--- of sequences of measures \cite{gibbs2002choosing}. For this reason we allow for flexibility in our analysis, and work with the family of $f$-divergences. These metrics have a convex function $f$ as a free parameter. We use several important members of this family as running examples: the Kullback-Leibler and the $\chi^2$ divergences, and the total variation and Hellinger distances. Previous non-asymptotic analyses of importance sampling  have focused on the first two. For instance, \cite{CP15}  suggested ---under certain concentration condition on the density--- the necessity and sufficiency of the sample size being larger than the exponential of the Kullback-Leibler divergence between target and proposal, and \cite{agapiou2015importance} proved the  sufficiency of the sample size being larger than the $\chi^2$ divergence for autonormalized importance sampling for bounded test functions. Indeed the function $U$ in the above argument is given by $U(N)=\log N$ when $D$ is the Kullback-Leibler divergence, and $U(N)=N-1$ for the $\chi^2$ divergence, in agreement with \cite{CP15} and complementing \cite{agapiou2015importance}.   The Kullback-Leibler divergence plays also a key role in the asymptotic analysis, since it provides the rate function of the large deviation principle for the empirical measure \cite{sanov1958probability}, and it also appears in the rate function for weighted empirical measures \cite{hult2012large}. 

A second step in formalizing our idea is to agree on how to interpret the statement that importance sampling is not successful. In this regard, moderate mean squared error seems too exacting as a necessary criteria, since this statistic may even be infinite while the method gives small errors with overwhelming probability. Moderate mean squared error seems more adequate as a {\em sufficient} than a {\em necessary} requirement. We propose, following \cite{CP15}, to consider the method unsuccessful when there are test functions for which importance sampling gives significant errors with large probability. Our main results, Theorems \ref{theorem} and Theorem \ref{theorem2}, show that when the sample size is not sufficiently large in terms of the $f$-divergence between the target and the proposal, then the method breaks down ---with high probability--- for either   $\phi \equiv 1$ or $\phi_f := f\circ(d \p/ d\Q)$. Note that the latter test function depends on the choice of $f$-divergence; for a given choice of $f$, the $\Q$-integrability of $\phi_f$ will naturally determine the class of test functions for which our upper bounds are meaningful.

We close the introduction with a brief literature review and an outline of this paper. Importance sampling is a standard tool in computational statistics \cite{liu2008monte}. It was first proposed as a variance reduction technique for standard Monte Carlo integration \cite{kahn1953methods}, and has been extensively used in the simulation of  rare events since \cite{siegmund1976importance}, where the interest lies in computing the expectation of a given test function $\phi$ (typically an indicator). Importance sampling has received recent interest as a building block of particle filters \cite{del2004feynman}, \cite{doucet2001introduction}. In this complementary range of applications, which motivates our presentation, the interest lies in approximating a measure, and computing expectations over a {\em class} of test functions \cite{del2004feynman}, \cite{agapiou2015importance}. $f$-divergences were introduced in \cite{csisz1963informationstheoretische}, \cite{csisz1967information} and \cite{ali1966general} as a generalization of the Kullback-Leibler divergence \cite{kullback1951information}. They have been widely studied in information theory, and an in-depth treatment is given in \cite{liese2007convex}.  A recent survey of bounds on $f$-divergences is \cite{sason2015bounds}. Finally, \cite{gibbs2002choosing} contains a brief and clear exposition of the relationships between probability metrics, sufficient for the purposes of this paper.

Section \ref{sec:setting} provides the necessary background on importance sampling. Section \ref{sec:distances} briefly reviews $f$-divergences, and some bounds on and between them are established. The main results are in Section \ref{sec:mainsec}. Examples are given in Section \ref{sec:examples}, and we conclude in Section \ref{sec:conclusion}.

Notation:
We let $g:= d\p/d\Q$ denote the Radon-Nikodym derivative of $\p$ with respect to $\Q.$ We denote measures that are not necessarily probabilities with Greek letters. For any measure $\nu$ in $(\X,\F)$ and measurable function $\phi:\X \to \R$ we set $\nu(\phi) := \int_\X \phi(x) \nu(dx).$ Randomness arising from sampling is indicated with a superscript: $N$ for the number of samples and $n$ for the indices of the samples. Vectors are denoted in bold face, and ${\bf u}:= \{1/N,\ldots, 1/N\}$ denotes the uniform probability vector.

\section{Importance Sampling Background} \label{sec:setting}
Given samples $\{v^n\}_{n=1}^N$  from $\Q$, importance sampling approximates $\p(\phi)$ as follows:
\begin{equation*}
\p(\phi) = \int_{\X} \phi(x) \p(dx) = \int_{\X} \phi(x) g(x) \Q(dx) \approx \frac{1}{N} \sum_{n=1}^N \phi(v^n)g(v^n) = \sum_{n=1}^N w^n \phi(v^n),
\end{equation*}
where $w^n := g(v^n)/N.$ Our presentation does not cover autonormalized importance sampling, but we expect that our paradigm could be generalized with additional effort.
Recalling the definition of the particle approximation measure in \eqref{eq:weightedmeasure}, the previous display can be rewritten as
\begin{equation*}
\p(\phi) \approx \pi^N(\phi).
\end{equation*}
The right-hand side is an unbiased estimator of $\p(\phi),$ and its mean squared error is given by
\begin{equation*}
\text{  MSE}\bigl( \pi^N(\phi) \bigr) = \frac{\Var_\Q(g\phi)}{N}.
\end{equation*}
As noted in the introduction, we argue that small mean squared error is sufficient but not necessary for successful importance sampling.
An important point for further developments is that $\pi^N$ is a random measure, but in general it is not a \emph{probability} measure since the weights $w^n$ typically do not add up to one. It is clear, however, that in the large $N$ asymptotic 
\begin{equation*}
\sum_{n=1}^N w^n = \frac{1}{N} \sum g(v^n) \approx 1,
\end{equation*}
and precise statements about the quality of the above approximation can be made under different assumptions on the moments of $g$ under $\Q.$

\section{The Family of $f$-divergences} \label{sec:distances}
In this section we introduce the family of $f$-divergences, and a slight generalization for atomic measures that need not integrate to one. The section closes with useful upper bounds for the analysis of importance sampling.

 Given a convex function $f$ with $f(1) = 0$, the $f$-divergence between $\p$ and $\Q$ is defined by 
\begin{equation*}
\df(\p \| \Q) := \int_{\X} f\circ g(x)\, \Q(dx) \equiv \Q(f\circ g),
\end{equation*}
where, recall, $g=d\p/d\Q.$
The assumptions on $f$ and Jensen's inequality ensure that these divergences are non-negative. However,  $\df$ does not constitute in general a distance in the space ${\cal P}(\X)$ of probability measures on $(\X, \F)$: it typically does not satisfy the triangle inequality or the requirement of symmetry, it takes the value $\infty$ if $f\circ g$ is not $\Q$-integrable, and it may need to be redefined when the first argument is not absolutely continuous with respect to the second. Examples are given in Table \ref{tableinfo}, where it is shown {\em a} choice of $f$ (the choice is in general not unique) that results in the Kullback-Leibler and the $\chi^2$ divergences, and the total variation and Hellinger distances.
 We spell out the definitions here, and some useful characterizations:
\begin{align*}
\dkl(\p\| \Q) &:= \Q\bigl(g\log(g)\bigr)  \equiv \p\bigl(\log(g)\bigr), \\
\dchi(\p\|  \Q) &  :=\Q\bigl( g^2 \bigr) - 1  \equiv \p(g) -1, \\
\dtv(\p,  \Q) &  := \Q(|1-g|) \equiv \sup_{A\subset \F} | \p(A) - \Q(A)| \in [0,1], \\
\dhell(\p ,  \Q)^2 &  :=  \Q \bigl( (\sqrt{g} -1)^2 \bigr) \in [0,2].
\end{align*}

While $\dtv$ and $\dhell$ can be shown to be distances in ${\cal P}(\X)$, $\dkl$ and $\dchi$ are not. In particular, these latter divergences fail to be symmetric, a feature that makes them appealing for the analysis of importance sampling. Indeed, the very formulation of the method is built on an asymmetric premise (the absolute continuity of $\p$ with respect to $\Q$). Moreover, it is well acknowledged that it is desirable that the proposal has heavier tails than the target ---again an asymmetric requirement.

We have already stressed that we are not interested in the mean squared error as a statistic to discard estimators. The next result is, however, instructive for comparison purposes. It gives necessary conditions on the sample size for bounded mean squared error over bounded test functions. Here and later we will drop the arguments of the divergences when no confusion may arise. 

\begin{proposition}\label{proposition}
Let $\phi \equiv 1$ be the constant function $1,$ and let $C>0.$  If $\text{MSE}(\pi^N(\phi))\le C,$ then
\begin{eqnarray*}
N& \ge C^{-1}\dchi, \quad \quad N &\ge C^{-1} (\exp(\dkl) - 1),  \\
N &\ge 4C^{-1}  \dtv^2, \quad \quad N &\ge C^{-1} \dhell^2.
\end{eqnarray*}
\begin{proof}
First note that, for $\phi \equiv 1$, 
\begin{equation*}
\text{MSE}\bigl(\pi^N(\phi)\bigr) = \frac{ \Var_\Q(g)}{N} = \frac{\Q(g^2) - 1}{N} = \frac{\dchi}{N}.
\end{equation*}
This gives the bound for $\dchi.$ The bounds for the other metrics follow from the general bounds 
\begin{equation*}
\dkl \le \log(1+ \dchi), \quad \dtv \le \frac{\sqrt{\dchi}}{2}, \quad \dhell \le \sqrt{\dchi},
\end{equation*}
which can be found, for instance, in \cite{gibbs2002choosing}.
\end{proof}
\end{proposition}
\begin{remark}\label{remark1}
Always $\dtv\le 1$ and $\dhell \le \sqrt 2.$ Therefore the mean squared error calculation above gives, in fact, no bounds for these distances unless $\sqrt{CN} <2,$ respectively $\sqrt{CN}<\sqrt 2$. On the other hand, the bounds for $\dkl$ and $\dchi$ are meaningful for any values of $C$ and $N$, since these divergences  could \emph{a priori} take arbitrarily large values.
\end{remark}

We now generalize the definition of $f$-divergence to atomic measures that need not be probabilities. Let ${\bf{p}}:=\{ p_1, \ldots, p_N\}$ and ${\bf{q}}:=\{ q_1, \ldots, q_N\}$ be vectors with $p_i\ge 0$, $q_i >0,$ $	1\le i\le N$. Given a convex function $f$ with $f(1) = 0$, the $f$-divergence between ${\bf p}$ and ${\bf q}$ is defined by  
\begin{equation*}
\df( {\bf p} \| {\bf q }) : = \sum_{i=1}^N q_i f\Bigl( \frac{p_i}{q_i} \Bigr).
\end{equation*}
This generalization will be useful for the analysis. We note, however, that the interpretation of these generalized $f$-divergences as ``distance" is somewhat lost, as they can take negative values. The following lemma gives a general upper bound on the $f$-divergence between arbitrary weights and uniform weights ${\bf u}$. Examples are given in Table \ref{tableinfo}.

\begin{table}
\begin{center}
\caption{\label{tableinfo}Summary of the four $f$-divergences used as examples in this paper. The third column contains the maximum value these divergences can take when the first argument is a probability vector and the second is the uniform probability vector ${\bf u}$. The fourth column is the generalization to the case where the first argument is a non-negative vector with total mass $\le 1+\epsilon.$ }
\begin{tabular}{l |  c  c c  c}
 Divergence & $f(x)$  &$U_f(N)$ & $U_f(N, \epsilon)$ \\ \hline
Kullback-Leibler & $x\log(x)$ & $\log(N)$ & $(1+\epsilon)\log\bigl\{ N (1+\epsilon)\bigr\}$\\
$\chi^2$ &  $(x-1)^2$ & $N-1$ & $N(1+\epsilon)^2-(1+2\epsilon)$ \\
Total variation  &  $|x-1|/2$ &$ 1-1/N$ & $1-1/N + \epsilon/2$ \\
Squared Hellinger  & $(\sqrt{x}-1)^2$   & $2(1-1/\sqrt{N})$ & $2\Bigl(1-\sqrt{\frac{1+\epsilon}{N}} + \epsilon/2\Bigr)$\\
\end{tabular}
\end{center}
\end{table}

\begin{lemma}\label{lemma}
Let ${\bf{p}}:=\{ p_1, \ldots, p_N\}$ and ${\bf u}=\{1/N, \ldots, 1/N\}$ be probability vectors. Then 
\begin{equation*}
\df( {\bf p} \| {\bf u } ) \le \frac{f(N) + (N-1)f(0)}{N}=: U_f(N).
\end{equation*}

If ${\bf p }$ has non-negative entries but it is allowed to have total mass $\sum p_i \le 1+\epsilon$, then 
\begin{equation}\label{eq:ufepsilon}
\df( {\bf p} \| {\bf u } ) \le \frac{f((1+\epsilon)N) + (N-1)f(0)}{N} =: U_f(N, \epsilon).
\end{equation}

Equality is achieved when $p_i=1$ (or $p_i = 1+\epsilon)$ for some $1\le i \le N.$

\begin{proof}
It follows from the convexity of $f$ that $\df(\, \cdot \, \| \,\cdot\,)$ is convex in its first argument (e.g. \cite{csiszar2004information}). Hence, by convexity, the ${\bf p}$ in the probability simplex that maximizes $\df({\bf p}\| {\bf u})$ is in one of the vertices of the simplex, i.e. there is $1\le i \le N$ with $p_i =1$. The expression for $U_f$ is then obtained by substituting such ${\bf p}$ in the definition of $\df.$ The proof when ${\bf p}$ is allowed to have total mass $1+\epsilon$ is identical.
\end{proof}
\end{lemma}

\section{Main Results. Necessary Sample Size}\label{sec:mainsec}
This section contains the main results of the paper, and formalizes the heuristic argument outlined in the introduction.
\begin{theorem}[Necessary sample size]\label{theorem}
Let $\epsilon, \delta>0$, and let $U_f(N,\epsilon)$ be defined as in \eqref{eq:ufepsilon}. Assume that with $\Q$-positive probability $i)$ and $ii)$ below hold simultaneously
\begin{itemize}
\item[i)] $\pi^N(1) \le  1+\epsilon.$ 
\item[ii)] $ | \Q(f\circ g) - \Q_{\text{\tiny MC}}^N(f\circ g)| \le \delta.$
\end{itemize}
Then
\begin{equation*}
\df(\p\| \Q) \le U_f(N,\epsilon) + \delta.
\end{equation*}
\begin{proof}
Note that
\begin{equation*}
\Q(f\circ g) = \df(\p\|\Q), \quad \quad \Q_{\text{\tiny MC}}^N(f \circ g) = D_f({\bf w}^N \| {\bf u}).
\end{equation*}
Let $A\in \F$  be the set where $i)$ and $ii)$ hold. Using Lemma \ref{lemma}
\begin{align*}
\Q(A) \df(\p\|\Q) &  \le \int_A \Bigl( \bigl|\df(\p\|\Q) - D_f({\bf w}^N \| {\bf u})\bigr| + \bigl|D_f({\bf w}^N \| {\bf u})\bigr|  \Bigl) d\Q\\
&=   \int_A \Bigl(  \bigl| \Q(f\circ g) - \Q_{\text{\tiny MC}}^N(f\circ g)\bigr| + \bigl|D_f({\bf w}^N \| {\bf u})\bigr|    \Bigl) d\Q \\
& \le  \Q(A) \bigl\{ \delta + U_f(N,\epsilon)\bigr\}.
\end{align*}
Since by assumption $\Q(A)>0$ the proof is complete.
\end{proof}
\end{theorem}

For the Kullback Leibler and the $\chi^2$ divergences, condition ii) in Theorem \ref{theorem} can be rewritten in terms of the particle measure $\pi^N.$ The next result is thus a reformulation of Theorem \ref{theorem}, where the necessary sample size is derived by using the expressions for   $U_f(N,\epsilon)$ in Table \ref{tableinfo}. The proof is immediate and will be omitted.

\begin{theorem}[Necessary sample size: Examples] \leavevmode \label{theorem2}
Let $\epsilon, \delta>0$.
\begin{enumerate}
\item If $N< (1+\epsilon)^{-1} \exp\Bigl(\frac{\dkl - \delta}{1+\epsilon} \Bigr),$ then, with probability at least $1/2,$ either
\begin{equation*}
\pi^N(1) - \p(1) > \epsilon \quad\quad \text{or} \quad\quad  | \pi^N(\log g) - \p(\log g)| > \delta.
\end{equation*}
\item If $N< (1+\epsilon)^{-2} (1+2\epsilon +\dchi - \delta),$ 
then, with probability at least $1/2,$ either 
\begin{equation*}
\pi^N(1) - \p(1) > \epsilon \quad\quad \text{or} \quad \quad |\pi^N( g) - \p(g)| > \delta.
\end{equation*}
\item If $N<(1+\epsilon/2 + \delta - \dtv)^{-1}$ 
then, with probability at least $1/2,$ either 
\begin{equation*}
\pi^N(1) - \p(1) > \epsilon \quad\quad \text{or} \quad \quad |\Q(|g-1|/2) - \Q_{\text{\tiny MC}}^N(|g-1|/2) | > \delta.
\end{equation*}
\item If $N<4(1+\epsilon) (2+\epsilon + \delta - \dhell^2)^{-2}$ 
then, with probability at least $1/2,$ either 
\begin{equation*}
\pi^N(1) - \p(1) > \epsilon \quad\quad \text{or} \quad \quad  |\Q\bigl((\sqrt g -1)^2\bigr) - \Q_{\text{\tiny MC}}^N\bigl((\sqrt g -1)^2\bigr) | > \delta.
\end{equation*}
\end{enumerate} 
\end{theorem}

 \begin{remark}
Note that $\epsilon$ and $\delta$ in Theorems \ref{theorem} and \ref{theorem2} are arbitrary. In particular, choosing $\delta^* \in (0,1)$ and replacing  $\delta$ by $\delta^* \Q(f\circ g)$ immediately gives relative error conditions, as opposed to the absolute ones above. It could also be interesting to consider  scaling $\epsilon$ and $\delta$ with $N$, but we do not pursue this here.
\end{remark}

\begin{remark}\label{remarkalternativeview}
Theorems \ref{theorem} and \ref{theorem2} can be viewed as yielding necessary conditions on the sample size $N$ for fixed $\p$ and $\Q$ or, alternatively, as giving necessary conditions on the $f$-divergence between $\p$ and $\Q$ for fixed $N$. Both interpretations are useful: in practice and depending on the problem it may be more convenient to guarantee that the necessary conditions are met by increasing the sample size or by reducing the $f$-divergence between target and proposal by means of a tempering scheme.
\end{remark}

\begin{remark}\label{remark2}
The bounds in the previous theorems, as opposed to those in Proposition \ref{proposition}, are derived without assuming finite mean squared error. Moreover,  when the required sample sizes above are not met, we show specific test functions for which the method gives significant error with large probability. The first two items in Theorem \ref{theorem2} are meaningful provided $\dkl$ and $\dchi$ are finite, that is, provided there is $\Q$-integrability of  $g\log g$ and $g^2$, respectively. 
 Thus  ---when these conditions hold--- the bounds give a necessary sample size for importance sampling over the classes of functions 
\begin{equation*}
\fkl := \{ \phi: \Q(\phi\log \phi)<\infty\}, \quad \quad \fchi:= \{\phi: \Q(\phi^2)<\infty\} \subset \fkl.
\end{equation*} 
  Whenever  the more restrictive condition $\Q(g^2)<\infty$ holds the analysis with $\dchi$ is sharper, since always $\dkl \le \log(1+\dchi)$. We informally summarize the above discussion as follows:
 \begin{itemize}
 \item If $\Q(g^2)<\infty,$ then $N\approx \dchi$ is required for accuracy over $\fchi.$ 
 \item if $\Q(g^2) = \infty$ but $\Q(g\log g)<\infty,$ then $N\approx \exp(\dkl)$ is required for accuracy over $\fkl.$ 
 \end{itemize}
The analysis with $\dkl$ is thus of interest when $\Q(g^2)=\infty.$ For instance, if $\Q(g^2)=\infty$  importance sampling has infinite mean square error for $\phi \equiv 1$ (see Proposition \ref{proposition}) but, as demonstrated in \cite{CP15}, the  $L^1$ error may be moderate if $N>\dkl.$ In such a case it is perhaps advisable to reconsider how to monitor the performance of the method. Precisely, letting 
 $$ \widehat{w}^n:= \frac{w^n}{\sum_{n=1}^N w^n}$$
 be the normalized weights, we suggest the use of
\begin{equation*}
\esskl :=\frac{N}{\exp\Bigl(\dkl({\bf \widehat{w}}^N \| {\bf u})\Bigr)} = \frac{N}{\exp \Bigl(\frac1N \sum_{n=1}^N \widehat{w}^n \log\bigl(N\widehat{w}^n\bigr)\Bigr)} \in [1,N],
\end{equation*}
rather than the usual
\begin{equation*}
\esschi := \frac{N}{1+\dchi({\bf \widehat{w}}^N \| {\bf u})} = \frac{1}{\sum_{n=1}^N (\widehat{w}^n)^2}\in [1,N],
\end{equation*}
 to monitor the effective sample size. 
 
 Finally, note that the bounds on $\dtv$ and $\dhell$ do not pose any restriction on the integrability of the density $g$. However, their sharpness is very limited. In particular the largest required sample size they can provide (achieved when $\dtv=1$ or $\dhell = \sqrt2$) is given, respectively,  by 
 \begin{align*}
 (\epsilon/2 + \delta)^{-1}, \quad \quad 4(1+\epsilon)/(\epsilon + \delta)^{-2}.
 \end{align*}
For $\dkl$ and $\dchi$ the required sample size grows without bound, regardless of $\epsilon$ and $\delta$, as the target and proposal become further apart.  This is in analogy with Remark \ref{remark1}.
 \end{remark}

\begin{table}
\begin{center}
\caption{\label{table2}Necessary sample size given by Theorem \ref{theorem2} for $\Q = N(0,1)$ and $\p = N(m, 1).$}
\begin{tabular}{l   c  c c  c c c}
 & $\nkl$ &$\nchi$ & $\ntv$ & $\nhell$ \\
$m=2$ & $5.11$  & $\approx 47$  & $\approx 2$  & $2.20$\\
$m=2.5$ &  $14.22$ & $\approx 350$ & $\approx 3$   &$3.53$ \\
$m=3$  & $49.63$  & $\approx 10^3$ & $\approx 4$  & $6.10$\\
$m=3.5$  & $217.45$    & $\approx 10^4$ &  $\approx 4$& $11.00$ \\
\end{tabular}
\end{center}
\end{table}

\begin{table}
\begin{center}
\caption{\label{table3}Necessary sample size given by Theorem \ref{theorem2} for $\Q = N(0,1)$ and $\p = N(0, \sigma^2).$}
\begin{tabular}{l  c  c c  c c c}
 & $\nkl$ &$\nchi$ & $\ntv$ & $\nhell$ \\
$\sigma^2=10^{-9}$ & $6.50 \times 10^3$  &  $\approx 10^4$  & $\approx 6$ & $94.39$ \\
$\sigma^2=10^{-4}$ & $34.67$  &$\approx 85$   &  $\approx 6$  &$18.87$  \\
$\sigma^2=16$  & $215.23$  & $---$  & $\approx 1$& $1.78$  \\
$\sigma^2=25$  & $1.05\times 10^4$    & $---$ &  $\approx 1$ & $2.12$ \\
\end{tabular}
\end{center}
\end{table}

\section{Examples}\label{sec:examples}
We illustrate the bounds in Theorem \ref{theorem2} with simple examples. In all  of them we let the proposal be a standard Gaussian distribution, $\Q = N(0,1),$ and we take as target a Gaussian distribution $\p = N(m,\sigma^2)$ with mean $m$ and covariance $\sigma^2$ to be specified later. In this framework  $\dkl$ and $\dhell$ can be computed in closed form, and we estimate $\dchi$ and $\dtv$ by an intensive Monte Carlo computation with $10^8$ samples. Clearly, absolute continuity of $\p$ with respect to $\Q$ always holds. We fix $\epsilon = \delta = 0.1$. Tables \ref{table2} and \ref{table3} contain the necessary sample size given by Theorem \ref{theorem2} for all four metrics $\dkl, \dchi, \dtv$ and $\dhell.$ In Table \ref{table2},  $\sigma^2=1$ is fixed, and we vary the values of $m$. For any value of $m$, $g$ has bounded $\Q$-moments of all orders.
In Table \ref{table3},  $m=0$ is fixed and we vary the values of $\sigma^2.$ Here, $g$ has finite $\Q$-moment of order $\alpha>0$ iff $\sigma^2 \le \alpha/(\alpha -1).$ In particular  $\dchi$ is finite iff $\sigma^2<2,$ and when this condition is not met the bound in Theorem \ref{theorem2} becomes meaningless (see Remark \ref{remark2}).

In order to compare the results derived with each divergence, it is important to keep in mind the discussion in Remark \ref{remark2}. Tables \ref{table2} and \ref{table3} show, as predicted, that $\dchi$ yields the largest required sample size. The analysis with this metric becomes, however, meaningless when $\Q(g^2)$ is infinity (Table \ref{table3}, $\sigma^2\ge 2$). Table \ref{table3} is illustrative. As mentioned above, it is desirable that the tails of the proposal are heavier than those of the target (i.e.  $\sigma^2 <1).$ The total variation and Hellinger distances are symmetric in their arguments, and hence they fail to see the problems arising from heavy target tails (large $\sigma^2$). The asymmetric divergences $\dkl$ and $\dchi$ do capture the asymmetric behavior of the problem: $\dchi$ in a dramatic fashion as it  becomes infinity for $\sigma^2 \ge 2,$ and $\dkl$ gives a bound of the same order  when the ratio of the covariances of $\Q$ and $\p$ is $10^9$ as when it is $1/25.$
 It is perhaps more surprising to see the poor bounds that these metrics yield in Table \ref{table2}, where target and proposal differ only by a shift, but this is also explained by the discussion in Remark \ref{remark2}.

\section{Conclusion} \label{sec:conclusion}
The approach and results in this paper give new insight into the fundamental challenge that importance sampling faces as a building block of more sophisticated algorithms: the target and the proposal must be sufficiently close. As noted elsewhere \cite{agapiou2015importance}, the often claimed curse of dimension of importance sampling \cite{BBL08}---and consequently of particle filters--- hinges exclusively on the observation that measures {\em tend to} become further apart in larger dimensional spaces. However, this needs not be the case, and is indeed not so in many data assimilation problems of applied interest \cite{agapiou2015importance}, \cite{chorin2013conditions}. 
Topics for further research include the extension to autonormalized importance sampling and other related algorithms, and the question of how to optimize over the choice of $f$ with a given $\Q$-integrability of $f\circ g$ to achieve the largest necessary condition on the required sample size.

\bibliographystyle{biometrika}
\bibliography{isbib}

\begin{thebibliography}{19}
\expandafter\ifx\csname natexlab\endcsname\relax\def\natexlab#1{#1}\fi

\bibitem[{Agapiou et~al.(2015)Agapiou, Papaspiliopoulos, Sanz-Alonso \&
  Stuart}]{agapiou2015importance}
\textsc{Agapiou, S.}, \textsc{Papaspiliopoulos, O.}, \textsc{Sanz-Alonso, D.}
  \& \textsc{Stuart, A.~M.} (2015).
\newblock Importance sampling: computational complexity and intrinsic
  dimension.
\newblock \textit{arXiv preprint arXiv:1511.06196} .

\bibitem[{Ali \& Silvey(1966)}]{ali1966general}
\textsc{Ali, S.~M.} \& \textsc{Silvey, S.~D.} (1966).
\newblock A general class of coefficients of divergence of one distribution
  from another.
\newblock \textit{Journal of the Royal Statistical Society. Series B
  (Methodological)} , 131--142.

\bibitem[{Bengtsson et~al.(2008)Bengtsson, Bickel, Li et~al.}]{BBL08}
\textsc{Bengtsson, T.}, \textsc{Bickel, P.}, \textsc{Li, B.} et~al. (2008).
\newblock Curse-of-dimensionality revisited: Collapse of the particle filter in
  very large scale systems.
\newblock In \textit{Probability and statistics: Essays in honor of David A.
  Freedman}. Institute of Mathematical Statistics, pp. 316--334.

\bibitem[{Chatterjee \& Diaconis(2015)}]{CP15}
\textsc{Chatterjee, S.} \& \textsc{Diaconis, P.} (2015).
\newblock The sample size required in importance sampling.
\newblock \textit{arXiv preprint arXiv:1511.01437} .

\bibitem[{Chorin \& Morzfeld(2013)}]{chorin2013conditions}
\textsc{Chorin, A.~J.} \& \textsc{Morzfeld, M.} (2013).
\newblock Conditions for successful data assimilation.
\newblock \textit{Journal of Geophysical Research: Atmospheres} \textbf{118},
  11--522.

\bibitem[{Csisz\'ar(1963)}]{csisz1963informationstheoretische}
\textsc{Csisz\'ar, I.} (1963).
\newblock Eine informationstheoretische ungleichung und ihre anwendung auf den
  beweis der ergodizit{\" a}t von markoffschen ketten.
\newblock \textit{Publ. Math. Inst. Hungar. Acad.} \textbf{8}, 95--108.

\bibitem[{Csisz\'ar(1967)}]{csisz1967information}
\textsc{Csisz\'ar, I.} (1967).
\newblock Information-type measures of difference of probability distributions
  and indirect observations.
\newblock \textit{Studia Sci. Math. Hungar.} \textbf{2}, 299--318.

\bibitem[{Csisz{\'a}r \& Shields(2004)}]{csiszar2004information}
\textsc{Csisz{\'a}r, I.} \& \textsc{Shields, P.~C.} (2004).
\newblock \textit{{Information theory and statistics: A tutorial}}.
\newblock Now Publishers Inc.

\bibitem[{Del~Moral(2004)}]{del2004feynman}
\textsc{Del~Moral, P.} (2004).
\newblock \textit{Feynman-Kac Formulae}.
\newblock Springer.

\bibitem[{Doucet et~al.(2001)Doucet, De~Freitas \&
  Gordon}]{doucet2001introduction}
\textsc{Doucet, A.}, \textsc{De~Freitas, N.} \& \textsc{Gordon, N.} (2001).
\newblock {An Introduction to Sequential Monte Carlo Methods}.
\newblock In \textit{{Sequential Monte Carlo Methods in Practice}}. Springer,
  pp. 3--14.

\bibitem[{Gibbs \& Su(2002)}]{gibbs2002choosing}
\textsc{Gibbs, A.~L.} \& \textsc{Su, F.~E.} (2002).
\newblock On choosing and bounding probability metrics.
\newblock \textit{International statistical review} \textbf{70}, 419--435.

\bibitem[{Hult \& Nyquist(2012)}]{hult2012large}
\textsc{Hult, H.} \& \textsc{Nyquist, P.} (2012).
\newblock Large deviations for weighted empirical measures arising in
  importance sampling.
\newblock \textit{arXiv preprint arXiv:1210.2251} .

\bibitem[{Kahn \& Marshall(1953)}]{kahn1953methods}
\textsc{Kahn, H.} \& \textsc{Marshall, A.~W.} (1953).
\newblock {Methods of reducing sample size in Monte Carlo computations}.
\newblock \textit{Journal of the Operations Research Society of America}
  \textbf{1}, 263--278.

\bibitem[{Kullback \& Leibler(1951)}]{kullback1951information}
\textsc{Kullback, S.} \& \textsc{Leibler, R.~A.} (1951).
\newblock On information and sufficiency.
\newblock \textit{The annals of mathematical statistics} \textbf{22}, 79--86.

\bibitem[{Liese \& Vajda(1987)}]{liese2007convex}
\textsc{Liese, F.} \& \textsc{Vajda, I.} (1987).
\newblock \textit{{Convex Statistical Distances}}, vol.~95.
\newblock Teubner-Texte Zur Mathematik.

\bibitem[{Liu(2008)}]{liu2008monte}
\textsc{Liu, J.~S.} (2008).
\newblock \textit{Monte Carlo Strategies in Scientific Computing}.
\newblock Springer Science \& Business Media.

\bibitem[{Sanov(1958)}]{sanov1958probability}
\textsc{Sanov, I.~N.} (1958).
\newblock \textit{On the probability of large deviations of random variables}.
\newblock United States Air Force, Office of Scientific Research.

\bibitem[{Sason \& Verd{\'u}(2015)}]{sason2015bounds}
\textsc{Sason, I.} \& \textsc{Verd{\'u}, S.} (2015).
\newblock Bounds among $ f $-divergences.
\newblock \textit{arXiv preprint arXiv:1508.00335} .

\bibitem[{Siegmund(1976)}]{siegmund1976importance}
\textsc{Siegmund, D.} (1976).
\newblock {Importance sampling in the Monte Carlo study of sequential tests}.
\newblock \textit{The Annals of Statistics} , 673--684.

\end{thebibliography}

\end{document}